# Super heavy particle production in high energy heavy ion collision

A.B.Kurepin

Institute for Nuclear Research RAS, Moscow

**Abstract**. The estimate based on the parton model is made on the rate of production of Super Heavy Particle ( SHP ) in subthreshold collision of heavy ions at LHC. For the one month run of lead-lead collision the yield of 16 TeV particle is of the order of 70 per year.

## 1. Introduction

In the Grand Unification Theory it is assumed that the first massive particles which are responsible for the symmetry breaking up to the symmetry of the Standard Model at the distances larger than $10^{-29}$ cm have masses of the order of $10^{12}$ TeV. In the simple version of this Theory next masses arise only at distances about $10^{-16}$ cm, where the particles with masses of the order of $10^2$ GeV should be considered. They are connected to the breaking of SU(2) symmetry, so called Higgs scalar bosons. Therefore rather artificial hypothesis was proposed on the existence of the " gauge dessert", i.e. no particles have masses between $0.2 – 10^{12}$ TeV. However some other possibility for high mass production is predicted by Large Extra Dimension theories, where our 3+1 dimensional world is considered as a " brane " in the higher dimensional space. Then the Planck mass can be brought down to a level of the order of 1 TeV. Then the Super Heavy Particles with masses of several TeV like a mini Black Holes could be produced if the extra dimension is equal to seven.

Also more complicated models has been proposed with multistep breaking of the symmetry at super high energy down to the symmetry of the Standard Model, where some new particles could be adopted with masses in the interval filling the " gauge dessert". One of the most interesting possibility provides the Theory based on new symmetry principle connecting the particles with different statistics, so called Supersymmetry Theory. New types of heavy quarks and leptons are introduced in this theory: s – quarks and s – leptons and corresponding supersymmetry partners. In that respect we can take into account new possibilities for producing the masses up to $10^3$ TeV with the LHC with heavy ions as an opportunity to search for new super heavy supersymmetry particles. We can expect rather small coupling between these new types of physical objects and particles of the ordinary world. Therefore, one can look for super heavy particles with large lifetime.

___________________________________________________________________________
e-mail: kurepin@inr.ru     1

The unique possibility of relativistic heavy ion collisions is the production with considerable probability the particles which are kinematically forbidden for nucleon-nucleon collisions with the same energy per nucleon. The essential increase of the "subthreshold" production cross section for nucleus-nucleus collision normalized per one nucleon in comparison to nucleon-nucleus collision was observed [ 1-5]. This increase can reach two order of magnitude and could not be explained by Fermi motion in the incident nucleus [ 9 ]. " Subthreshold" particle production could be due to multinucleon or multiquark correlation in heavy nuclei.

The total energy in the center of mass system for Pb – Pb collisions at the LHC is about 1150 TeV. Therefore, the unique possibility could be considered to produce Super Heavy Particles ( SHP ) by nucleus – nucleus collisions with masses much larger than the center of mass energy in p – p collisions.

## 2. Parton model for particle production in heavy ion collision

For the analysis of the subthreshold hadron production at intermediate energies [ 1-5 ] the phenomenological parton model was proposed [ 6 ]. It was observed that the subthreshold and near threshold production cross section in proton – nucleus and nucleus collisions for pions, kaons and antiprotons could be expressed as a universal function of the scaling quark – parton parameters, like Bjorken scaling parameter $x$ for deep inelastic electron scattering but take into account the production of massive particles and now different in incident ( Z ) and in target ( X ) nuclei. The values Z and X larger than 1 arise to provide the production of particles at subthreshold energies for nucleon – nucleon process. As usually a physical interpretation of these scaling parameters in a quark-parton model means that X $P_b$ correspond to the parton 4 – momentum in a target nuclei, where $P_b$ is a 4 – momentum of a nucleon in a target nucleus. X could take the values in an interval 0 <X<A, where A is the atomic number of a target. On the other hand the same parton model could be applied to the incident particle Z $Pa$. .

For the antiproton production the value of X for fixed Z could be obtained:

$$X = \frac{Z(P_a P_d) + Z m_a m_n + \frac{1}{2}(m_n^2 - m_d^2)}{Z(P_a P_b) - Z m_a m_b - (P_b P_d) - m_b m_n} \qquad (1)$$

which is derived from the conservation of the 4- momentum in the collision. Here $P_d$ is the 4- momentum of the antiproton, $m_a$ $m_b$ $m_n$-mass of the nucleon in the incident and in the target nuclei, $m_d$ - mass of the antiproton. Also we assume that the total energy of SHP is much larger than the total energy of partons after the interaction.

For proton – nucleus and for nucleus – nucleus interactions the experimental data in a wide energy interval for different incident and target nucleon could be approximated with the same scaling law. Those are the data for antiproton production with Ne and Ni nuclei with 1 – 2 GeV/nucl. at GSI [ 8 ], proton and carbon nuclei with 3.65 GeV/nucl. at JINR [ 1, 2 ], with deuteron beam at KEK [ 7 ] and with protons and Si- beam at LBL for antiproton and K⁻ production [ 9 ]. The scaling on Fig.1 is observed with Z- parameter equal 1 for proton beam, 1.3 for deuteron, 2 for carbon beam and 3 for Si, Ne, and Ni nuclei [ 6 ]. The production cross sections were reduced by $(A_1 A_2)^{0.43}$ according to A- dependence investigated in [ 10 ] and due to the different absorption nuclear effects for pions, kaons and antiprotons.

The considerable increase of the ratio of kaon to pion production rates in nucleus – nucleus collision compared to the same ratio in deuteron – nucleus collision was also explained by the scaling dependence on Z parameter for Z < 1, but with the parameter X= 2 for incident carbon nucleus and X=1.5 for incident deuterons [ 11 ].

The considerable increase of the production rates for kaons and antiprotons compared to pions in nucleus – nucleus collisions was interpreted as the indication of collective parton effect and was quantitatively reproduced by introduction of the scaling parameters larger than 1. From the Fig.2 we see that the dependence of the Lorenz invariant inclusive cross section, or of the structure function, on the scaling parameter is close to the exponent. The "subthreshold" structure function on Fig. 2 could be approximated as:

$$(A_1 A_2)^{-0.43} \cdot E_1 \frac{d^3\sigma}{dp^3} \; [\text{mb GeV}^{-2} \; c^3 \; \text{sr}^{-1}] = 0.57 \exp(-X/0.158) \qquad (2)$$

where $A_1$ and $A_2$ are the mass numbers of colliding nuclei.

This curve can be used to describe the experimental data in the range of cross section $10^{-1} – 10^{-5}$ mb with the accuracy less than an order of magnitude for X = 0.2 – 3 for the data from about 2 to 400 GeV.

The enhancement of kaon and antiproton production is quantitatively reproduced by the introduction of large Z scaling parameter: Z = 1.3 for deuterons, Z = 3 for all heaver colliding nuclei. For incident protons Z = 1. To justify the application of this scaling dependence of the invariant inclusive cross section ultra relativistic energies we stress that the same scaling law was found for 8.9 GeV [ 12 ] and 400 GeV [ 13 ] protons incident on nuclei.

This scaling was observed for different quark flavors. Therefore one can expect that the model could be applied to some heavier unknown quarks and that these collective phenomena could be due to some general space – time properties of nucleus- nucleus interaction or to the universal structure function of a nucleus.

## 3. Estimate of production yield of Super Heavy Particle production

From equation ( 2 ) we see that the cross section strongly depends on the parameter X . Therefore to produce SHP with considerable probability one should look for the kinematics with smallest X and Z possible. To perform the estimation of SHP production cross section for LHC using the scaling dependence of structure function ( 2 ) for M >>$m_n$ we can determine Z and X from the relation on the threshold for SHP production:

$$\sqrt{S} = 2 E \sqrt{X \cdot Z} = M, \qquad (3)$$

where M is the mass of SHP, $m_n$ - the mass of nucleon, E is the energy per nucleon of colliding nuclei.

To estimate the production rate of low energy SHP with the mass M, the equation ( 2 ) could be used with Z= 3 for nucleus – nucleus collision and X to be determined from ( 3 ).

The absolute value of the momentum P of SHP is obtained from the equation

$$X = \frac{2Z(E \cdot E_1 - E \cdot P \cdot \cos\theta) - M^2}{4Z \cdot E^2 - 2(E \cdot E_1 + E \cdot P \cdot \cos\theta)} \qquad (4)$$

which is derived from the conservation of the 4- momentum in the collision. Here $E_1$ is the total energy of SHP, θ - angle between SHP and Z parton.

Assuming that SHP are produced with small Pt in the narrow forward – backward cone we get for the masses M larger than 1 TeV the velocities of SHP are of the order of beta = 0.1 – 0.9.

Accepting that the multi quark correlations are the intrinsic property of nuclei and does not depend on the collision energy, we can try to use the "subthreshold" structure function ( 2 ) to estimate the production rate of SHP depending on their mass in Pb - Pb collisions:

$$E_1 \frac{d^3\sigma}{dp^3} \ [\text{mb GeV}^{-2} \ c^3 \ \text{sr}^{-1}] = 56 \exp(-X/0.158) \qquad (5)$$

For the obtained luminosity L = 2 · $10^{27}$ cm$^{-2}$ sec$^{-1}$ for one month of the run and the opening angle of the forward detector about 100 msr we get the estimate of the SHP yield for the mass about 16 TeV near threshold with of the order of 70 per year. For larger SHP masses the production cross section is greatly reduced due to the increasing parameter X in the exponential dependence.

The question remains about the possibility of applicability of the empirical scaling law, obtained at intermediate energies, to interactions at ultrahigh LHC energies. Unfortunately, it is impossible to measure the scattering and production of particles at angles greater than 90 degrees on the collider experimental facilities , that is, at X> 1.

Very promising possibilities for studying subthreshold phenomena are opened up in the case of experiments with a fixed target at the LHC [ 14 - 16 ]. The AFTER or ALICE-FT project provides for the placement of a solid or gaseous target in the beam halo when using all detectors of the ALICE facility. In this case, it is possible to measure the production of particles, in particular antiprotons, at large parameters X [ 17 ] even at X > 1, which will make it possible to check the existence of the scaling, considered in this work, at the energies of the LHC ion beams. For a fixed target with a beam energy of 2.76 TeV per nucleon and Z = 3 at the antiproton production threshold, from formula (4) we obtain X =1.15 and larger X values for production angles greater than 90 degrees.

## 4. Conclusion

The extrapolation of the scaling observed in the parton model for subthreshold particle production in the collision of heavy ions to high energies was made, in order to obtain estimates of the cross section and yield for the production of particles with masses up to 16 TeV. It is proposed to investigate subthreshold phenomena in an experiment with a fixed target at the LHC.

This research was supported by the RFBR/CNRS grant 18-52-15007.

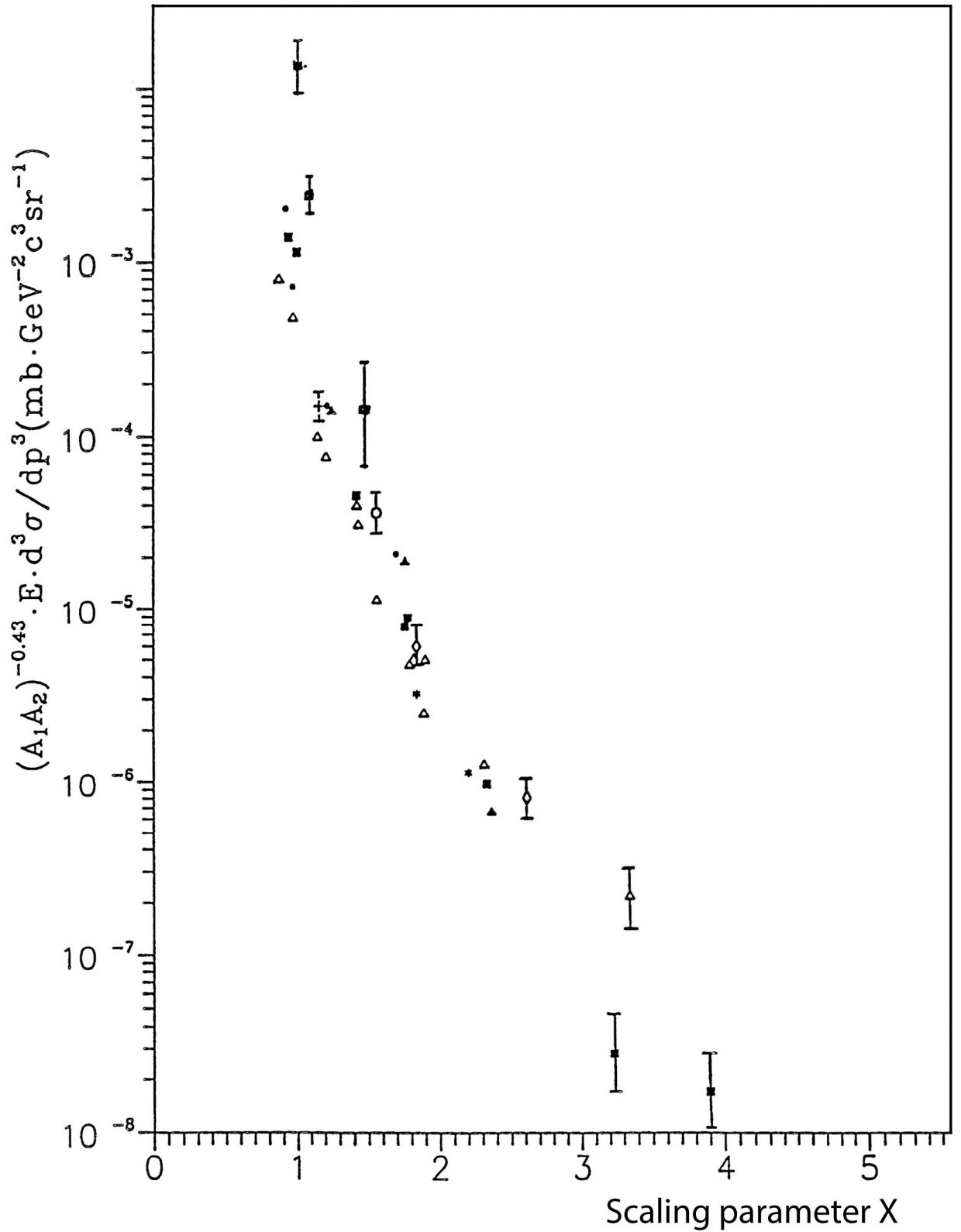

Fig. 1. Systematics of Lorentz invariant antiproton subthreshold production cross section dependence on the scaling parameter X with Z = 1 for incident protons, Z = 1.3 for deuterons, Z = 2 for carbon ions, Z = 3 for heavy nuclei. Points p + C [ 1,2 ], open circles d + C [ 1,2 ], crosses C + C and C + Cu [ 1,2 ], triangles p + C and d + C [ 7 ], squares p + Cu [ 9 ], rhombs Si + Si [ 9 ], stares Ne + Sn and Ni + Ni [ 8 ].

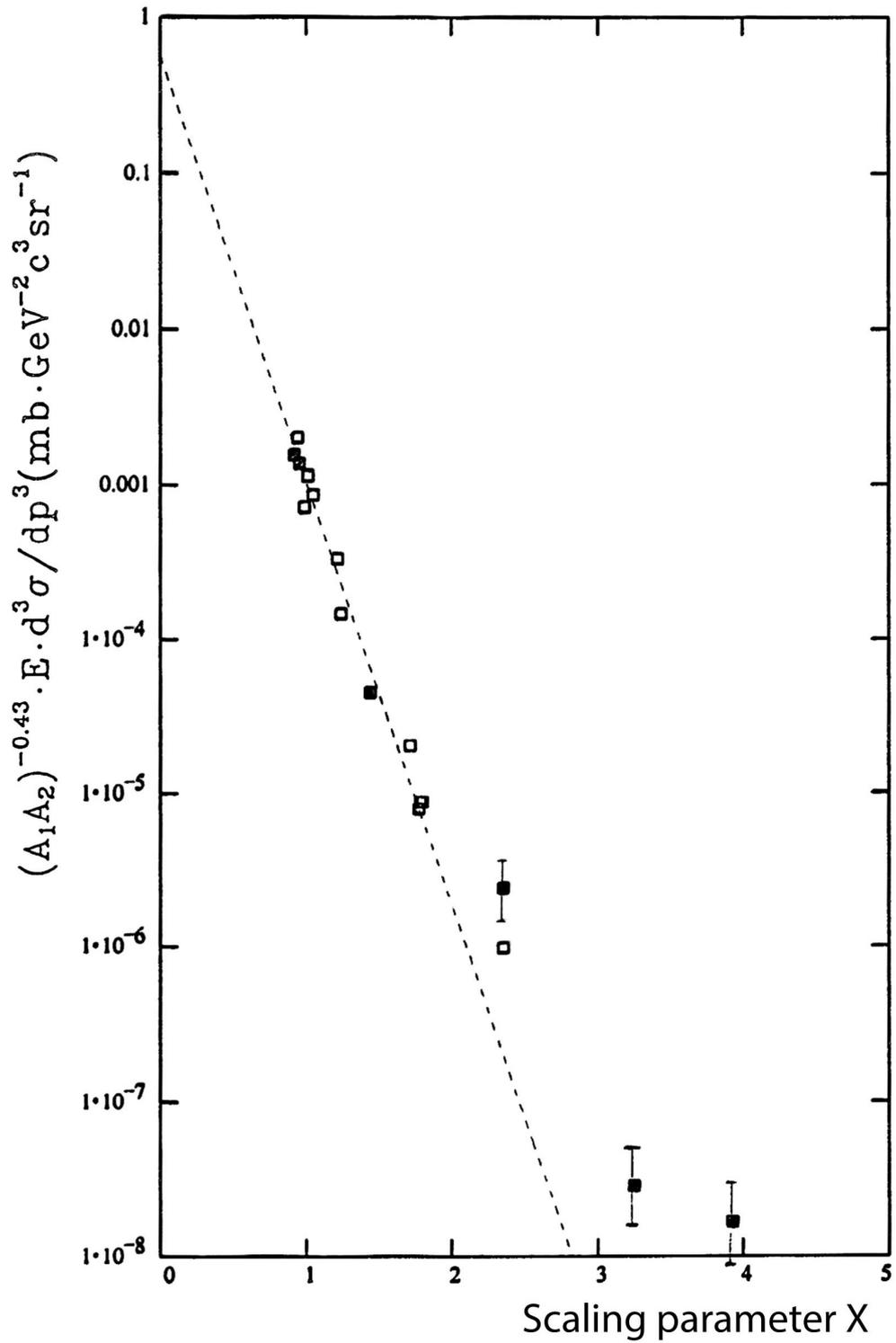

Fig. 2. $\chi^2$ – fit to the invariant antiproton production cross section dependence on X- scaling parameter. Only data for the interaction of protons with carbon ( open marks ) and with copper (filled marks ) are shown